\documentclass[onecolumn,10pt]{article}
\usepackage{graphicx}
\usepackage{epstopdf}
\usepackage{amsfonts}
\usepackage{times}
\usepackage{cancel}
\usepackage{soul}
\newcommand{\ket}[1]{| #1 \rangle}

\begin{document}

\title{Quantum Correlations \\ and the Measurement Problem}

\author{Jeffrey Bub\\ \small \textit{Philosophy Department and Institute for Physical Science and Technology}\\  \small \textit{University of Maryland, College Park, MD 20742, USA}}
\date{}
\normalsize

\maketitle

\begin{abstract}
The transition from classical to quantum mechanics rests on the recognition that the structure of information is not what we thought it was: there are operational, i.e., phenomenal, probabilistic correlations that lie outside the polytope of local correlations. Such correlations  cannot be simulated with classical resources, which generate classical correlations represented by the points in a simplex, where the vertices of the simplex represent joint deterministic states that are the common causes of the correlations. The `no go' hidden variable theorems tell us that we can't shoe-horn correlations outside the local polytope into a classical simplex by supposing that something has been left out of the story. The replacement of the classical simplex by the quantum convex set as the structure representing probabilistic correlations is the analogue for quantum mechanics of the replacement of Newton's Euclidean space and time by Minkowski spacetime in special relativity. The nonclassical features of quantum mechanics, including the irreducible information loss on measurement, are generic features of correlations that lie outside the local correlation polytope. This paper is an elaboration of these ideas, and its consequences for the measurement problem of quantum mechanics. A large part of the difficulty is removed by seeing that the inconsistency in reconciling the entangled state at the end of a quantum measurement process with the definiteness of the macroscopic pointer reading and the definiteness of the correlated value of the measured micro-observable is only apparent and depends on a stipulation that is not required by the structure of the quantum possibility space. Replacing this stipulation by an alternative consistent stipulation resolves the problem.

\end{abstract}

\section{Introduction}
\label{intro}
What is quantum mechanics about? In the case of special relativity, the answer is clear: special relativity is \emph{about the structure of space-time}. The replacement of Newton's Euclidean space and time with a different geometric structure, Minkowski space-time, in which space and time are relative to the state of motion of a system, followed from the recognition of a property of light, that `there is never any overtaking of light by light in empty space,' as Hermann Bondi puts it \cite[p. 27]{Bondi1967}, together with a revised relativity principle, that `velocity doesn't matter for physics' \cite[p. 20]{Bondi1967}, i.e., for electromagnetic as well as mechanical phenomena. 

A variety of answers have been proposed for quantum mechanics: that quantum mechanics is about energy being quantized in discrete lumps or quanta, or that it is about particles that are wavelike, or that it is about the universe, as we perceive it, continually splitting into countless co-existing quasi-classical universes, with many copies of ourselves, and so on. A more mundane answer is implicit in the theory of quantum information: quantum mechanics is about probabilistic correlations, i.e., \emph{about the structure of information}, insofar as a theory of information is essentially a theory of probabilistic correlations. We live in a universe in which there are correlations that lie outside the polytope of local correlations. Such correlations cannot be simulated with classical resources, which generate classical correlations represented by the points in a simplex, where the vertices of the simplex represent joint deterministic states that are the common causes of the correlations. The replacement of the classical simplex by the quantum convex set as the structure representing probabilistic correlations is the analogue for quantum mechanics of the replacement of Newton's Euclidean space and time by Minkowski spacetime in special relativity. 

The following discussion is an elaboration of this idea, and its consequences for the measurement problem of quantum mechanics. \S 1 is an introduction to  classical, quantum, and superquantum correlations, and the differences between them. \S 2 considers the question: if operational or phenomenal probabilistic correlations can lie outside the local correlation polytope and hence cannot be simulated with classical resources,  what principle constrains these correlations to the convex set of quantum correlations? Why not superquantum correlations? \S 3 is a discussion of two recent results, by Pusey, Barrett, and Rudolph \cite{PBR2012}, and by Colbeck and Renner \cite{ColbeckRenner2012}, about the interpretation of quantum states. The issue is framed as the question of whether quantum states are `epistemic' or `ontic.' As we'll see, the significance of these new results is more properly seen in relation to the Bell \cite{BellEPR} and Kochen and Specker \cite{KochenSpecker} `no go' theorems about possible extensions or completions of quantum mechanics, traditionally  formulated as the question of whether `hidden' variables underly the probability distributions defined by quantum states. \S 4 proposes a solution to the measurement problem as it arises in this information-theoretic view of quantum mechanics. 

\section{Correlations}
\label{sec:corr}

Consider the simple case of measurements of  two binary-valued observables, $x \in \{0,1\}$ with outcomes $a \in \{0,1\}$, performed by Alice in a region $\mathbf{A}$, and $y \in\{0,1\}$ with outcomes $b \in \{0,1\}$, performed by Bob in a separated region $\mathbf{B}$. Correlations are expressed by a correlation array of joint probabilties as in Table \ref{tab:correlationarray}. The probability $p(00|00)$ is to be read as $p(a=0,b=0|x=0,y=0)$, i.e., as a joint conditional probability, and the probability $p(01|10)$ is to be read as  $p(a=0,b=1|x=1,y=0)$, etc. (I drop the commas for ease of reading; the first two slots in $p(--|--)$ before the conditionalization sign $|$ represent the two possible measurement outcomes for Alice and Bob, respectively, and the second two slots after the conditionalization sign represent the two possible observables that Alice and Bob choose to measure, respectively.) 
\begin{table}[h!]
\begin{center}
\begin{tabular}{|ll|ll|ll|}
\hline
   &$x$&$0$ & &$1$&\\
   $y$&&&&&\\\hline
  $0$ &&$p(00|00)$&$ p(10|00)$  & $p(00|10)$&$ p(10|10)$     \\
   &&$p(01|00)$&$p(11|00)$  & $p(01|10)$&$ p(11|10)$  \\
\hline
   $1$&&$p(00|01)$&$ p(10|01)$  & $p(00|11)$&$ p(10|11)$   \\
  &&$p(01|01)$&$ p(11|01)$  & $p(01|11)$&$ p(11|11)$   \\
\hline
\end{tabular}
\end{center}
  \caption{Correlation array}
  \label{tab:correlationarray}
\end{table}

There are four probability constraints: the sum of the probabilities in each square cell of the array in Table \ref{tab:correlationarray} is $1$, since the sum is over all possible outcomes, given the two observables that are measured. The marginal probability of $0$ for Alice or  for Bob, for a given pair of observables, is obtained by adding the probabilities in the left column of each cell or the top row of each cell, respectively, and the marginal probability of $1$  for Alice or  for Bob, for a given pair of observables, by adding the probabilities in the right column of each cell or the bottom row of each cell, respectively. The measurement outcomes are uncorrelated if the joint probability is expressible as a product of marginal or local probabilities for Alice and Bob; otherwise they are correlated. 

Now consider all possible correlation arrays of the above form. They  represent the pure and mixed states of a bipartite system with two binary-valued observables for each subsystem and form a regular convex polytope with 256 vertices, where the vertices represent the extremal deterministic arrays or pure states with probabilities $0$ or $1$ only, e.g., the array in Table \ref{tab:signalingcorr}.\footnote{A regular polytope is the multi-dimensional analogue of a regular polygon, e.g., an equilateral triangle, or a square, or a pentagon in two dimensions. A convex set is, roughly, a set such that from any point in the interior it is possible to `see' any point on the boundary.} The polytope is the closed convex hull of the vertices, i.e., the smallest closed convex set containing the vertices. There are four possible arrangements of $0$'s and $1$'s that add to $1$ in each square cell of the correlation array (i.e., one $1$ and three $0$'s), and four cells, hence $4^{4} = 256$ vertices. The 16 probability variables in the correlation array are constrained by the four probability constraints. It follows that the 256-vertex polytope is 12-dimensional.\footnote{Thanks to Tony Sudbury for clarifying this (private communication).}  A general correlation array  is represented by a point in this polytope, so the probabilities in the array can be expressed (in general, non-uniquely) as convex combinations of the $0, 1$ probabilities in extremal correlation arrays (just as the probability of one of two alternatives, $0$ or $1$, can be represented as a point on a line between the points $0$ and $1$ because it can be expressed as a convex combination of the extremal end points).

The correlation array in Table \ref{tab:signalingcorr} defines a set of correlations that allow instantaneous signaling between Alice and Bob. Think of the $x$-values and $a$-values as Alice's inputs and outputs, respectively, and similarly for Bob with respect to the $y$-values and $b$-values. So the two Alice-inputs ($x = 0$ or $x=1$) correspond to the two Alice-observables, and the two Bob-inputs ($y=0$ or $y=1$) correspond to the two Bob-observables, and each observable can take two values, $0$ or $1$. In Table \ref{tab:signalingcorr} , Alice's output is the same as Bob's input. Similarly, Bob's output is the same as Alice's input. So an input by Alice or Bob is instantaneously revealed in  a remote output. There are 240 similar signaling extremal deterministic correlation arrays in the total set of 256 extremal deterministic correlation arrays. The remaining 16 extremal deterministic correlation arrays are no-signaling arrays.
\begin{table}[h!]
\begin{center}
\begin{tabular}{|ll|ll|ll|} \hline
   &$x$&$0$ & &$1$&\\
   $y$&&&&&\\\hline
  $0$ &&$p(00|00) = 1$&$ p(10|00) = 0$  & $p(00|10) = 0$&$ p(10|10) = 0$     \\
   &&$p(01|00) = 0$&$p(11|00) = 0$  & $p(01|10)=1$&$ p(11|10) = 0$  \\\hline
   $1$&&$p(00|01)=0$& $p(10|01)=1$  & $p(00|11)=0$&$ p(10|11)=0$   \\
  &&$p(01|01)=0$&$ p(11|01)=0$  & $p(01|11)=0$&$ p(11|11)=1$   \\\hline
\end{tabular}
\end{center}
  \caption{Extremal signaling deterministic correlation array}
  \label{tab:signalingcorr}
\end{table}

The no-signaling principle can be formulated as follows:  no information should be available in the marginal probabilities of outputs in region $\mathbf{A}$ about alternative choices made by Bob in region $\mathbf{B}$, i.e., Alice, in region $\mathbf{A}$ should not be able to tell what Bob measured in region $\mathbf{B}$, or whether Bob performed any measurement at all, by looking at the statistics of her measurement outcomes, and conversely.  Formally:
 \begin{eqnarray}
\sum_{b}p(a,b|x,y) \equiv p(a|x,y) & = & p(a|x), y \in \{0,1\} \label{eqn:nosignal1} \\
\sum_{a}p(a,b|x,y) \equiv p(b|x,y) & = & p(b|y), x \in \{0,1\} \label{eqn:nosignal2} 
\end{eqnarray}
Here $p(a,b|x,y)$ is the probability of obtaining the pair of outputs $a, b$ for the pair of inputs $x,y$. The probability $p(a|x,y)$ is the marginal probability of obtaining the output $a$ for $x$  when Bob's input is $y$, and $p(b|x,y)$ is the marginal probability of obtaining the output $b$ for $y$ when Alice's input is $x$. The no-signaling principle requires Alice's marginal probability  $p(a|x,y)$ to be independent of Bob's choice of input in region $\mathbf{B}$ (and independent of whether there was any input in region $\mathbf{B}$ at all), i.e.,  $p(a|x,y) = p(a|x)$, and similarly for Bob's marginal probability $p(b|x,y)$ with respect to Alice's inputs: $p(b|x,y) = p(b|y)$. 

Note that the no-signaling principle is simply a constraint on the marginal probabilities, not a relativistic constraint on the motion of a physical entity through space-time \emph{per se}---there is no reference to $c$, the velocity of light. What is excluded is that something happening \emph{here} has an \emph{immediate} effect over \emph{there}. Communication between Alice and Bob involves the exchange of messages encoded as physical signals that move between them through space at a certain velocity. If this constraint were violated, then instantaneous (and hence superluminal) signaling would be possible.

The joint probabilities in the 16 no-signaling deterministic correlation arrays can all be expressed as products of marginal or local probabilities for Alice and Bob separately. For example, the deterministic correlation array in which the outputs are both $0$ for all possible input combinations, as in Table \ref{tab:nonsignalingcorr}, is a no-signaling array and the joint probabilities can be expressed as a product of local probabilities: a marginal Alice-probability of $1$ for the output $0$ given any input, and a marginal Bob-probability of $1$ for the output $0$ given any input. This is, of course, not the case for the 240 signaling deterministic correlation arrays.
 \begin{table}[h!]
 \begin{center}
\begin{tabular}{|ll|ll|ll|} \hline
   &$x$&$0$ & &$1$&\\
   $y$&&&&&\\\hline
  $0$ &&$p(00|00) = 1$&$ p(10|00) = 0$  & $p(00|10) = 1$&$ p(10|10) = 0$     \\
   &&$p(01|00) = 0$&$p(11|00) = 0$  & $p(01|10)=0$&$ p(11|10) = 0$  \\\hline
   $1$&&$p(00|01)=1$& $p(10|01)=0$  & $p(00|11)=1$&$ p(10|11)=0$   \\
  &&$p(01|01)=0$&$ p(11|01)=0$  & $p(01|11)=0$&$ p(11|11)=0$   \\\hline
\end{tabular}
\end{center}
\caption{Extremal no-signaling deterministic correlation array}
  \label{tab:nonsignalingcorr}
\end{table}

Now suppose the correlations are as in Table \ref{tab:prcorr}. These correlations define  a Popescu-Rohrlich box (PR-box), a hypothetical device considered by Popescu and Rohrlich \cite{PopescuRohrlich94} to bring out the difference between classical, quantum, and superquantum no-signaling correlations. 
\begin{table}[h!]
 \begin{center}
\begin{tabular}{|ll|ll|ll|} \hline
   &$x$&$0$ & &$1$&\\
   $y$&&&&&\\\hline
  $0$ &&$p(00|00) = 1/2$&$ p(10|00) = 0$  & $p(00|10) = 1/2$&$ p(10|10) = 0$     \\
   &&$p(01|00) = 0$&$p(11|00) = 1/2$  & $p(01|10)=0$&$ p(11|10) = 1/2$  \\\hline
   $1$&&$p(00|01)=1/2$&$ p(10|01)=0$  & $p(00|11)=0$&$ p(10|11)=1/2$   \\
  &&$p(01|01)=0$&$ p(11|01)=1/2$  & $p(01|11)=1/2$&$ p(11|11)=0$   \\\hline
\end{tabular}
\end{center}
\caption{PR-box correlation array}
  \label{tab:prcorr}
  \end{table}

PR-box correlations can be defined as follows:
\begin{equation}
a\oplus b = x\cdot y \label{eqn:PRbox}
\end{equation}
where $\oplus$ is addition mod 2, i.e., 
\begin{itemize}
\item[$\bullet$] same outputs (i.e., 00 or 11) if the inputs are 00 or 01 or 10 
\item[$\bullet$] different outputs (i.e., 01 or 10) if  the inputs are 11
\end{itemize}
with the assumption that the marginal probabilities are all $1/2$ to ensure no signaling, so the outputs $00$ and $11$ are obtained with equal probability when the inputs are not both $1$, and the outputs $01$ and $10$ are obtained with equal probability when the inputs are both $1$. 

A PR-box functions in such a way that if Alice inputs a $0$ or a $1$, her output is $0$ or $1$ with probability $1/2$, irrespective of Bob's input, and irrespective of whether Bob inputs anything at all; similarly for Bob. The requirement is simply that whenever there are in fact two inputs, the inputs and outputs are correlated according to (\ref{eqn:PRbox}). 

A PR-box can function only once, so to get the statistics for many pairs of inputs one has to use many PR-boxes. In this respect, a PR-box mimics a quantum system: after a system has responded to a measurement (produced an output for an  input), the system is no longer in the same quantum state, and one has to use many systems prepared in the same quantum state to exhibit the probabilities associated with a given quantum state.

The 16 vertices defined by the local no-signaling deterministic states are the vertices of a polytope: the polytope of local correlations. The local correlation polytope is included in a no-signaling nonlocal polytope, defined by the 16 vertices of the local polytope together with an additional 8 nonlocal vertices, one of these nonlocal vertices representing the standard PR-box as defined above, and the other seven vertices representing  PR-boxes obtained from the standard PR-box by local reversible operations (relabeling the $x$-inputs, and the $a$-outputs conditionally on the $x$-inputs, and the $y$-inputs, and the  $b$-outputs conditionally on the $y$-inputs). For example, the correlations in Table \ref{tab:trprcorr} define a PR-box. Note that the 16 vertices of the local polytope can all be obtained from the vertex represented by Table \ref{tab:nonsignalingcorr} by similar local reversible operations. Both the 16-vertex local polytope and the 24-vertex no-signaling nonlocal polytope are 8-dimensional: in addition to the four probability constraints (one for each cell in the correlation array), there are four no-signaling constraints.\footnote{Bell's locality conditions, which characterize the local polytope, are tighter no-signaling constraints: no-signaling conditional on the values of local hidden variables.}
 \begin{table}[h!]
\begin{center}
\begin{tabular}{|ll|ll|ll|} \hline
   &$x$&$0$ & &$1$&\\
   $y$&&&&&\\\hline
  $0$ &&$p(00|00) = 0$&$ p(10|00) = 1/2$  & $p(00|10) = 0$&$ p(10|10) = 1/2$     \\
   &&$p(01|00) = 1/2$&$p(11|00) = 0$  & $p(01|10)=1/2$&$ p(11|10) = 0$  \\\hline
   $1$&&$p(00|01)=0$&$ p(10|01)=1/2$  & $p(00|11)=1/2$&$ p(10|11)=0$   \\
  &&$p(01|01)=1/2$&$ p(11|01)=0$  & $p(01|11)=0$&$ p(11|11)=1/2$   \\\hline
\end{tabular}
\end{center}
 \caption{Locally transformed PR-box correlation array (relative to Table 4)}
  \label{tab:trprcorr}
\end{table}

Correlations represented by points in the local polytope can be simulated with classical resources, which generate classical correlations represented by the points in a simplex, where the vertices represent joint deterministic states (extremal states) that are the common causes of the correlations. Probability distributions over these extremal states---mixed states---are represented by points in the interior or boundary of the simplex. A simplex is a regular convex polytope generated by $n+1$ vertices that are not confined to any $(n-1)$-dimensional subspace, e.g., a tetrahedron as opposed to a square. The lattice of subspaces of a simplex (the lattice of vertices, edges, and faces) is a Boolean algebra, with a 1-1 correspondence between the vertices, corresponding to the atoms of the Boolean algebra, and the facets (the $(n-1)$-dimensional faces), which correspond to the co-atoms. The classical simplex---in this case a 16-vertex  simplex---represents the correlation polytope of probabilistic states of a bipartite classical system with two binary-valued observables for each subsystem; the associated Boolean algebra represents the classical possibility structure. 

Suppose Alice and Bob are allowed certain resources. What is the optimal probability that they can perfectly simulate the correlations of a PR-box?

In units where $a = \pm 1, b = \pm 1$,\footnote{It is convenient to change units here to relate the probability to the usual expression for the Clauser-Horne-Shimony-Holt correlation, where the expectation values are expressed in terms of $\pm 1$ values for $x$ and $y$ (corresponding to the relevant observables). Note that `outputs same' or `outputs different' mean the same thing whatever the units, so the probabilities $p(\mbox{outputs same}|xy)$ and $p(\mbox{outputs different}|xy)$ take the same values whatever the units, but the expectation value $\langle xy\rangle$ depends on the units for $x$ and $y$.}
\begin{equation}
\langle 00\rangle = p(\mbox{outputs same}|00) - p(\mbox{outputs different}|00)
\end{equation}
so:
\begin{eqnarray}
p(\mbox{outputs same}|00) & = & \frac{1 +  \langle 00\rangle}{2} \\
p(\mbox{outputs different}|00) & = & \frac{1-\langle 00\rangle}{2}
\end{eqnarray}
and similarly for input pairs 01, 10, 11. 

It follows that the probability of successfully simulating a PR-box is given by:
\begin{eqnarray}
p\mbox{(successful sim)} & = & \frac{1}{4}(p(\mbox{outputs same}|00) + p(\mbox{outputs same}|01) + \nonumber \\
& &  p(\mbox{outputs same}|10) + p(\mbox{outputs different}|11)) \\
& = & \frac{1}{2}(1 + \frac{K}{4}) = \frac{1}{2}(1 + E)
\end{eqnarray}
where $K = \langle 00\rangle + \langle 01\rangle + \langle 10\rangle - \langle 11\rangle$ is the Clauser-Horne-Shimony-Holt (CHSH) correlation. 

Bell's locality argument \cite{BellEPR} in the Clauser-Horne-Shimony-Holt version \cite{CHSH} shows that if Alice and Bob are limited to classical resources, i.e., if they are required to reproduce the correlations on the basis of shared randomness or common causes established before they separate (after which no communication is allowed), then $|K_{C}| \leq 2$, i.e., $|E| \leq 
\frac{1}{2}$, so  the optimal probability of successfully simulating a PR-box is $\frac{1}{2}(1+\frac{1}{2}) = \frac{3}{4}$. 

If Alice and Bob are allowed to base their strategy on shared entangled states prepared before they separate, then the Tsirelson bound for quantum correlations requires that $|K_{Q}| \leq 2\sqrt{2}$, i.e., $|E| \leq \frac{1}{\sqrt{2}}$, so the optimal probability of successful simulation limited by quantum resources is $\frac{1}{2}(1+\frac{1}{\sqrt{2}}) \approx .85$. 

I use the term `nonlocal box' to refer to any no-signaling device with a probability array such that, using the nonlocal box, it is possible to successfully simulate a PR-box (extremal nonlocal box) with a probability greater than the classical value of $3/4$. We live in a nonlocal box world: a pair of qubits in an entangled quantum state constitutes a nonlocal box for certain pairs of measurements. As Popescu and Rohrlich observe, relativistic causality does not rule out simulating a PR-box with a probability greater than $\frac{1}{2}(1+\frac{1}{\sqrt{2}})$: there are possible worlds described by  superquantum theories that allow nonlocal boxes with no-signaling correlations  stronger than quantum correlations, in the sense that $\frac{1}{\sqrt{2}} < E \leq 1$. The correlations of a PR-box saturate the CHSH inequality ($E=1$), and so represent a limiting case of no-signaling correlations.

\begin{figure*}
  \includegraphics[width=0.75\textwidth]{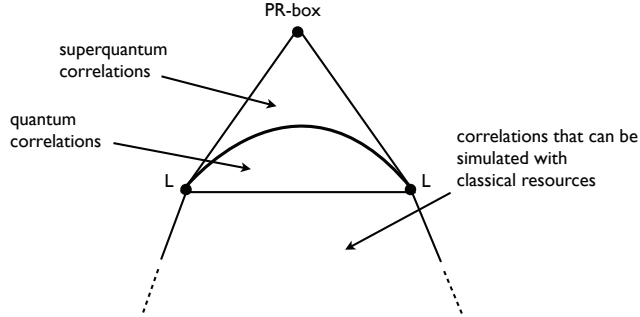}
\caption{Schematic representation of part of the space of no-signaling correlations for a bipartite system with binary input and output. The vertices L are the deterministic vertices of the local no-signaling polytope. Correlations in this polytope can be simulated with classical resources. The superquantum polytope is bounded by the sixteen vertices L of the local polytope together with eight PR-box vertices, which represent the strongest possible nonlocal correlations consistent with the no-signaling principle. The quantum convex set is bounded by a continuous set of vertices and is not a polytope. It lies between the 16-vertex local polytope and the 24-vertex superquantum no-signaling nonlocal polytope.} 
\label{fig:nosigcorr}
\end{figure*}

For two binary-valued observables of a bipartite quantum system, the correlations form a spherical convex set that  is not a polytope, with a continuous set of extremal points between the 16-vertex local correlation polytope characterized by Bell's inequalities (or the CHSH inequalities) and the 24-vertex no-signaling nonlocal polytope, which is itself included in the 256-vertex nonlocal polytope with 240 vertices that represent deterministic signaling states. By Bell's theorem, only correlations in the 16-vertex local polytope can be classically simulated by points in a simplex, where the vertices of the simplex represent joint deterministic states that are the common causes of the correlations. See Fig. \ref{fig:nosigcorr}.

A simplex has the rather special property that a mixed state, represented by a point in the interior of the simplex, can be expressed \emph{uniquely} as a mixture (convex combination) of extremal or pure states, the vertices of the simplex. \emph{No other polytope or convex set has this feature.} So in the class of no-signaling theories, classical (= simplex) theories are rather special. If there is no unique decomposition of mixed states into deterministic pure states for the operational or phenomenal probabilities, as is the case for the local 16-vertex polytope, then either there is some restriction on access to the full information encoded in the pure states, i.e., there is something left out of the story (which could, in principle, be added to complete the story as a simplex theory in which the vertices represent the common causes of correlations), or the correlations are outside the local polytope and there is no theoretical account that provides an explanation of correlations in terms of deterministic pure states without violating the no-signaling principle: any no-signaling explanation of correlations will have to include indeterministic states like PR-box states. 

To illustrate this non-uniqueness for the local correlation polytope, which is not a simplex, denote a deterministic state by the sequence $--;--$, where the first two slots denote the two possible Alice-outputs for the two possible Alice-inputs, and the second two slots after the semi-colon denote the two possible Bob-outputs for the two possible Bob-inputs. The equal-weight mixture, $m_{1}$, of the four deterministic states:
\begin{description}
\item $00;00$
\item $11;11$
\item $00;11$
\item $11;00$
\end{description}
is equivalent to the equal-weight mixture, $m_{2}$, of the four different deterministic states:
\begin{description}
\item $01;01$
\item $10;10$
\item $01;10$
\item $10;01$
\end{description}

 \begin{table}[h!]
\begin{center}
\begin{tabular}{|ll|ll|ll|} \hline
   &$x$&$0$ & &$1$&\\
   $y$&&&&&\\\hline
  $0$ &&$p(00|00) = 1/4$&$ p(10|00) = 1/4$  & $p(00|10) = 1/4$&$ p(10|10) = 1/4$     \\
   &&$p(01|00) = 1/4$&$p(11|00) = 1/4$  & $p(01|10)=1/4$&$ p(11|10) = 1/4$  \\\hline
   $1$&&$p(00|01)=1/4$&$ p(10|01)=1/4$  & $p(00|11)=1/4$&$ p(10|11)=1/4$   \\
  &&$p(01|01)=1/4$&$ p(11|01)=1/4$  & $p(01|11)=1/4$&$ p(11|11)=1/4$   \\\hline
\end{tabular}
\end{center}
 \caption{Correlation array associated with the two equivalent mixtures $m_{1}$ and $m_{2}$}
  \label{tab:equiv}
\end{table}
Both mixtures correspond to the same point in the local correlation polytope, represented by the array in Table \ref{tab:equiv}. For each mixture, the probabilities of the four possible Alice-Bob outputs, $00,01,10,11$, are equal for any Alice-input and Bob-input. Also, for each mixture, the marginal probabilities of Alice's two possible outputs, $0$ or $1$, are both $1/2$, for any input, and similarly for Bob. So the two mixtures are equivalent for Alice and for Bob, separately, assuming they each have access to only one input and associated output at a time, and they are also equivalent for Alice and Bob jointly, assuming they have access to only one pair of inputs and associated outputs at a time. The mixtures can only be distinguished in a  theory that assumes access to both inputs and outputs at the same time, for Alice and for Bob. This is the case in a theory in which the 16 deterministic states are represented by the vertices of a simplex, where the representation of mixtures, represented by points in the simplex, in terms of pure states, represented by the vertices, is unique. A 16-vertex simplex is a 15-dimensional polytope: there is only one probability constraint, that the sum of the sixteen probabilities $p(--|--)$ should sum to 1. The (operational) local correlation polytope characterizes a situation where access to the information encoded in the simplex is restricted.

If the operational or phenomenal correlations lie outside the local correlation polytope, there there is no theoretical explanation of the correlations in terms of a simplex theory, if we demand no violation of the no-signaling principle. Any theoretical explanation of such correlations will have to involve indeterministic pure states, such as PR-boxes, and the decomposition of mixed states into pure states will be non-unique. For such theories, there can be no general cloning procedure capable of copying an arbitrary extremal state without violating the no-signaling principle, and so \emph{there can be no measurement in the non-disturbing sense available in classical theories}, where it is in principle possible, via measurement, to extract sufficient information about an extremal state to produce a copy of the state without irreversibly changing the state. For a nonlocal box theory, \emph{there is  a necessary information loss on measurement}.

To see this, consider a PR-box and suppose there is some parameter $\lambda$ that Bob is able to measure, and   that Bob could somehow extract sufficient information via this measurement to know the outputs of both possible inputs to his part of the PR-box. Suppose Bob's information is that the output values for his two possible inputs are $j$ and $k$.

Consider a reference frame in which Alice's input occurs before Bob's measurement of $\lambda$.  If $j=k$, then Bob can infer that Alice's input was $0$, because otherwise, to satisfy the PR-box constraint (\ref{eqn:PRbox}), if Alice's input was 1, Alice's output would have to be the same as Bob's output value for  his input $0$, but different from Bob's output value for his input $1$, which is impossible if Bob's output values for his two inputs are the same. Similarly, if $j \neq k$, then Bob can infer that Alice's input was $1$, because otherwise, if Alice's input was $0$, Bob's output values for his two inputs would have to be the same as Alice's output for her input $0$, which would require Bob's output values to be the same. So the no-signaling principle is violated: Alice can signal instantaneously to Bob.

Allen Stairs has pointed out\footnote{Private communication.} that this conclusion follows only if Alice's output does not depend on Bob's input, e.g., as in Table \ref{tab:context1} for the case $j=k$ and as in  Table \ref{tab:context2} for the case $j \neq k$. But if Alice's output depends on Bob's input, then Bob can signal instantaneously to Alice, which again  involves a violation of the no-signaling principle. 
\begin{table}[h!]
\begin{center}
\begin{tabular}{|ll|ll|ll|} \hline
   &$x$&$0$ & &$1$&\\
   $y$&&&&&\\\hline
  $0$ &&$p(00|00) = 1$&$ p(10|00) = 0$  & $p(00|10) = 1$&$ p(10|10) = 0$     \\
   &&$p(01|00) = 0$&$p(11|00) = 0$  & $p(01|10)=0$&$ p(11|10) = 0$  \\\hline
   $1$&&$p(00|01)=1$&$ p(10|01)=0$  & $p(00|11)=0$&$ p(10|11)=1$   \\
  &&$p(01|01)=0$&$ p(11|01)=0$  & $p(01|11)=0$&$ p(11|11)=0$   \\\hline
\end{tabular}
\end{center}
 \caption{Bob's output is $0$ for all input pairs; Alice's output is $0$ for the input pairs $00, 01, 10$, but $1$ for the input pair $11$, i.e., Alice's output is nonlocally contextual and depends both on her input and on Bob's input.}
  \label{tab:context1}

\end{table}

 \begin{table}[h!]
 \begin{center}
\begin{tabular}{|ll|ll|ll|} \hline
   &$x$&$0$ & &$1$&\\
   $y$&&&&&\\\hline
  $0$ &&$p(00|00) = 1$&$ p(10|00) = 0$  & $p(00|10) = 1$&$ p(10|10) = 0$     \\
   &&$p(01|00) = 0$&$p(11|00) = 0$  & $p(01|10)=0$&$ p(11|10) = 0$  \\\hline
   $1$&&$p(00|01)=1$&$ p(10|01)=0$  & $p(00|11)=0$&$ p(10|11)=1$   \\
  &&$p(01|01)=0$&$ p(11|01)=0$  & $p(01|11)=0$&$ p(11|11)=0$   \\\hline
\end{tabular}
\end{center}
\caption{Bob's output  is the same as his input; Alice's output  is $0$ for the input pairs $00, 10, 11$, but $1$ for the input pair $01$, i.e., Alice's output is nonlocally contextual and depends both on her input and on Bob's input.}
  \label{tab:context2}
\end{table}

Now consider a reference frame in which Bob measures $\lambda$ before Alice's input. Then there is a constraint on Alice's choice of input: her input is $0$ if Bob's output values for his inputs are the same, and her input is $1$ if Bob's output values for his inputs are different. Again (Stairs' clarification), this follows only if Alice's output does not depend on Bob's input. But if Alice's output is nonlocally contextual and depends on Bob's input, then there is a violation of the no-signaling principle rather than a constraint on Alice's freedom to choose her input. So either there is a violation of `free choice,' i.e., what is sometimes called
 `superdeterminism'---Alice's input choice is not free but is is determined (or, more generally, constrained) by the value of  $\lambda$---or there is a violation of the no-signaling principle.

Of course, the order of Alice's and Bob's inputs could depend on the reference frame, but this does not alter the conclusion: if we reject the possibility of instantaneous signaling, or we assume that Alice's and Bob's input choices are unconstrained or  free, in the sense that these choices are independent of any information that is in principle available before the choices, then there must be a necessary information loss about the output value for input $0$ if Bob obtains information about the output value for input $1$, and conversely, and similarly for Alice.

For similar reasons, Bob can't clone his part of a PR-box. Suppose he could. If Bob inputs 0 into his part of a PR-box and 1 into the cloned part, then he could infer Alice's input from his two outputs if Alice's input occurs before his inputs: same outputs indicate that Alice's input was $0$; different outputs indicate that Alice's input was $1$. So Alice could signal instantaneously to Bob.  If Alice's input occurs after Bob's inputs, then Alice's input choice depends on whether Bob's output values are the same or different. Again (Stairs' clarification), this conclusion follows only if Alice's output does not depend on Bob's input. So if we assume that there is no constraint on Alice's freedom to choose her input, or that there is no violation of the no-signaling principle, then cloning part of a PR-box is impossible.

\section{Why Quantum Correlations?}
\label{sec:why}

The quantum theory is a nonlocal box theory, i.e., it is a no-signaling theory with counter-intuitive probabilistic features like those of a PR-box. Hilbert space as a projective geometry
(i.e., the subspace structure of Hilbert space) represents the structure of the space of possibilities and determines the kinematic part of quantum mechanics. This includes the association of  Hermitian
operators with observables, the Born probabilities, the von Neumann-L\"{u}ders
conditionalization rule, and the unitarity constraint on the dynamics, which is related to the possibility structure via a theorem of Wigner \cite{Wigner1959},\cite{Uhlhorn1963}. The possibility space is a non-Boolean space in which there are built-in, structural probabilistic constraints on correlations between events (associated with the angles between the rays representing extremal events)---just as in special
relativity the geometry of Minkowski space-time represents spatio-temporal
constraints on events. These are kinematic, i.e., pre-dynamic, objective probabilistic or information-theoretic constraints
on events to which a quantum dynamics of matter and fields conforms, through its symmetries, just
as the structure of  Minkowski space-time imposes spatio-temporal kinematic constraints on events to which a relativistic dynamics conforms.

The transition from classical to relativistic physics resolves a conflict between a contingent property of light, that there is no overtaking of light by light, and the Euclidean structure of Newtonian space and time.  It was Einstein's genius to see that treating the Galilean relativity principle as a structural principle applying to all of physics, i.e., to electromagnetic as well as dynamical phenomena (so the laws of physics are required to be the same in different reference frames moving at constant relative velocity) yields Minkowski spacetime---to put it somewhat anachronistically, since of course it was Minkowski who later codified Einstein's insight in terms of a specific spacetime structure. 

Schematically:
\begin{itemize}
\item[$\bullet$] relativity principle (`velocity doesn't matter'), a structural principle applying to all of physics
\item[$\bullet$]  light postulate (`no overtaking of light by light'), a contingent property of light
\item[$\bullet$]  \st{Newtonian spacetime} Minkowski spacetime
\end{itemize}

Here is Lorentz's reaction, from the conclusion of the 1916 edition of \emph{The Theory of Electrons and its Applications to the Phenomena of Light and Radiant Heat} \cite{Lorentz1916}:
\begin{quotation}
I cannot speak here of the many highly interesting applications which Einstein has made of this principle [of relativity]. His results concerning electromagnetic and optical phenomena \ldots agree in the main with those which we have obtained in the preceding pages, the chief difference being that Einstein simply postulates what we have deduced, with some difficulty and not altogether satisfactorily, from the fundamental equations of the electromagnetic field. By doing so, he may certainly take credit for making us see in the negative result of experiments like those of Michelson, Rayleigh and Brace, not a fortuitous compensation of opposing effects, but the manifestation of a general and fundamental principle.

Yet, I think, something may also be claimed in favour of the form in which I have presented the theory. I cannot but regard the aether, which can be the seat of an electromagnetic field with its energy and its vibrations, as endowed with a certain degree of substantiality, however different it may be from all ordinary matter. In this line of thought, it seems natural not to assume at starting that it can never make any difference whether a body moves through the aether or not, and to measure distances and lengths of time by means of rods and clocks having a fixed position relative to the aether.
\end{quotation}

The analogous contingent fact for the quantum revolution is the existence of nonlocal entanglement; specifically, that there are operational or phenomenal correlations outside the local polytope.  This requires extending the classical simplex structure for probabilistic correlations to the quantum convex set. The no-signaling principle (a structural constraint) requires that probabilistic correlations lie inside the no-signaling polytope. But as Popescu and Rohrlich \cite{PopescuRohrlich94} first pointed out, the quantum convex set is a proper subset of the no-signaling polytope---the no-signaling principle by itself does not constrain probabilistic correlations to the quantum convex set. What is the structural principle that constrains correlations to the quantum convex set inside the no-signaling polytope?

\begin{itemize}
\item[$\bullet$]  ?
\item[$\bullet$]  nonlocal entanglement (there are operational correlations outside the local polytope), a contingent fact
\item[$\bullet$]  \st{classical information: simplex structure of probabilistic correlations} quantum information: quantum convex set for probabilistic correlations
\end{itemize}

There are various proposals in the literature for the ?-principle. The most promising candidate, in my view,  is the principle of \emph{information causality} proposed by Pawlowski et al \cite{Pawlowski+2009}. Information causality is a generalization of the no-signaling principle. It can be interpreted as a principle characterizing system separability, or a limitation of what Bohr referred to as quantum `wholeness.'

Information causality says that Bob's information gain about a data set of Alice (previously unknown to him), on the basis of his local resources (which may be correlated with Alice's local resources) and a single use by Alice of an information channel with classical capacity $m$, is bounded by the classical capacity of the channel. For $m=0$, this is equivalent to no signaling. Zukowski calls the principle `causal information access.'\footnote{Private communication}  The proposal is that quantum mechanics optimizes causal information access. 

Here is a simple way to see the significance of information causality as a constraint. If Bob has to guess the value of any designated one of $N$ bits held by Alice, and Alice can send Bob just one bit of information, then Bob can do better exploiting quantum correlations than classical correlations (shared randomness). Information causality sets a limit on how much better he can do. 

Suppose the probability of Bob guessing the $k$'th bit correctly is $P_{k}$. The binary entropy of $P_{k}$ is defined as:
\begin{equation}
h(P_{k}) = - P_{k}\log{P_{k}} - (1-P_{k})\log{(1-P_{k})}
\end{equation}

If Bob knows the value of the bit he has to guess, $P_{k} = 1$, so $h(P_{k})= 0$. If Bob has no information about the bit he has to guess, $P_{k} = 1/2$, i.e., his guess is at chance, and $h(P_{k})= 1$. So $h(P_{k})$ varies between $0$ and $1$.

If Alice sends Bob one classical bit of information, information causality requires that Bob's information about the $N$ unknown bits increases by at most one bit. If the bits in Alice's list are unbiased and independently distributed, Bob's information about an arbitrary bit $b = k$ in the list cannot increase by more than $1/N$ bits, i.e., for Bob's guess about an arbitrary bit in Alice's list, the binary entropy $h(P_{k})$ is at most $1/N$ closer to 0 from the chance value 1: $h(P_{k})\geq 1 - 1/N$.

So information causality is violated when $h(P_{k}) < 1 - 1/N$ or, taking $N = 2^{n}$, the condition for a violation of information causality is $h(P_{k}) < 1 - \frac{1}{2^{n}}$. Since it can be shown that $P_{k} = \frac{1}{2}(1+E^{n})$ \cite{Pawlowski+2009,Bub2012}, we have a violation of information causality when $h(\frac{1}{2}(1+E^{n})) < 1 - \frac{1}{2^{n}}$.

For classical correlations, $E = \frac{1}{2}$, so in the case $n=1$ (where Bob has to guess one of two bits), $P_{k} = \frac{3}{4}$, and:
\[
h(P_{k}) \approx .81
\]
For quantum correlations, $E = \frac{1}{\sqrt{2}}$, so for $n=1, P_{k} = \frac{1}{2}(1+\frac{1}{\sqrt{2}}) \approx .85$, and:
\[
h(P_{k}) \approx .60
\]
For PR-box correlations, $E=1$, so for all $n$, $P_{k }=1$, and:
\[
h(P_{k}) = 0
\]

It can be shown that if $E \leq E_{T} = \frac{1}{\sqrt{2}}$, information causality is satisfied \cite{Pawlowski+2009,Bub2012}, i.e., 
\begin{equation}
h(\frac{1}{2}(1+E^{n}_{T})) \geq 1 = \frac{1}{2^{n}}, \mbox{for any } n
\end{equation}

If $E > E_{T}$, information causality is violated. 

If $E$ is very close to the Tsirelson bound $E_{T} = \frac{1}{\sqrt{2}} \approx .707$, then $n$ must be very large for a violation of information causality. For example, if $n=10$ and $E = .708$, then $h(P_{k}) \approx .99938$. There is no violation of information causality because $.99938 > 1 - \frac{1}{1024} \approx .9990$. In fact, we require $n \geq 432$ for a violation of information causality \cite{Bub2012}.

Quantum and classical theories satisfy information causality, but the question of whether information causality by itself is  sufficient to precisely delimit the quantum convex set  is still open. In the very special case of a bipartite system with two binary-valued observables, Allcock et al \cite{Allcock+2009} have been able to  exclude all but a very small part of the superquantum region on the basis of  information causality.

It is rather striking how closely Lorentz's reluctance to accept the theory of relativity is paralleled by Einstein's reluctance to accept the significance of the quantum revolution \cite{Einstein1948}:
\begin{quotation}
If one asks what, irrespective of quantum mechanics, is characteristic of the world of ideas of physics, one is first of all struck by the following: the concepts of physics relate to a real outside world, that is, ideas are established relating to things such as bodies, fields, etc., which claim a `real existence' that is independent of the perceiving subject---ideas which, on the other hand, have been brought into as secure a relationship as possible with sense-data. It is further characteristic of these physical objects that they are thought of as arranged in a physical space-time continuum.  An essential aspect of this arrangement of things in physics is that they lay claim, at a certain time, to an existence independent of one another, provided that these objects `are situated in different parts of space.' Unless one makes this kind of assumption about the independence of the existence (the `being-thus') of objects which are far apart from one another in space---which stems in the first place from everyday thinking---physical thinking in the familiar sense would not be possible. It is also hard to see any way of formulating and testing the laws of physics unless one makes a clear distinction of this kind.
\end{quotation}

\section{The Interpretation of Quantum States}
\label{sec:interp}

Schr\"{o}dinger  introduced the term `entanglement' to describe the non-simplex probabilistic
correlations of quantum mechanics, specifically the correlations of entangled Einstein-Podolsky-Rosen (EPR) states (quantum states of the sort considered by Einstein, Podolsky, and Rosen \cite{EPR} in their  argument for the incompleteness of quantum theory). He regarded entanglement as
\cite[p. 555]{Schr1}: `\textit{the} characteristic trait of
quantum mechanics, the one that enforces its entire departure from
classical lines of thought.' From this perspective (endorsed here), the transition from classical to quantum mechanics is the transition from a classical simplex theory  to a non-simplex theory of  operational or phenomenal probabilistic correlations that lie outside the local polytope. The conceptual problems of the theory reflect a clash between the structure of the associated possibility space and the standard ontology that goes along with classical (simplex) correlations. As far as these  problems are concerned, we might as well consider the class of superquantum theories rather than quantum mechanics: the conceptual problems are the same.

The existence of operational or phenomenal correlations outside the local correlation polytope means that we have to give up Einstein's view that physical objects have a `being-thus' (`So-sein' in German), in the sense that a physical object is characterized by definite properties prior to measurement that the observer is able to ascertain via measurement.

Recently, a new result by Pusey, Barrett, and Rudolph (PBR) \cite{PBR2012} has created a stir in the quantum foundations community. In a blog post by Matthew Leifer dated November 20, 2011 (http://mattleifer.info/2011/11/20/can-the-quantum-state-be-interpreted-statistically/), the issue is laid out particularly clearly:
\begin{quotation}
We have seen that there are two clear notions of state in classical mechanics: ontic states (phase space points) and epistemic states (probability distributions over the ontic states). In quantum theory, we have a different notion of state---the wavefunction---and the question is: should we think of it as an ontic state (more like a phase space point), an epistemic state (more like a probability distribution), or something else entirely?

Here are three possible answers to this question:
\begin{enumerate}
\item[1.] Wavefunctions are epistemic and there is some underlying ontic state. Quantum mechanics is the statistical theory of these ontic states in analogy with Liouville mechanics.
\item[2.] Wavefunctions are epistemic, but there is no deeper underlying reality.
\item[3.] Wavefunctions are ontic (there may also be additional ontic degrees of freedom, which is an important distinction but not relevant to the present discussion).
\end{enumerate}

I will call options 1 and 2 psi-epistemic and option 3 psi-ontic. Advocates of option 3 are called psi-ontologists, in an intentional pun coined by Chris Granade. Options 1 and 3 share a conviction of \emph{scientific realism}, which is the idea that there must be some description of what is going on in reality that is independent of our knowledge of it. Option 2 is broadly anti-realist, although there can be some subtleties here.

The theorem in the paper attempts to rule out option 1, which would mean that scientific realists should become psi-ontologists. I am pretty sure that no theorem on Earth could rule out option 2, so that is always a refuge for psi-epistemicists, at least if their psi-epistemic conviction is stronger than their realist one.

I would classify the Copenhagen interpretation, as represented by Niels Bohr, under option 2. One of his famous quotes is:
\begin{quote}
There is no quantum world. There is only an abstract physical description. It is wrong to think that the task of physics is to find out how nature is. Physics concerns what we can say about nature \ldots
\end{quote}
and `what we can say' certainly seems to imply that we are talking about our knowledge of reality rather than reality itself. Various contemporary neo-Copenhagen approaches also fall under this option, e.g. the Quantum Bayesianism of Carlton Caves, Chris Fuchs and Ruediger Schack; Anton Zeilinger's idea that quantum physics is only about information; and the view presently advocated by the philosopher Jeff Bub. These views are safe from refutation by the PBR theorem, although one may debate whether they are desirable on other grounds, e.g. the accusation of instrumentalism.

Pretty much all of the well-developed interpretations that take a realist stance fall under option 3, so they are in the psi-ontic camp. This includes the Everett/many-worlds interpretation, de Broglie-Bohm theory, and spontaneous collapse models. Advocates of these approaches are likely to rejoice at the PBR result, as it apparently rules out their only realist competition, and they are unlikely to regard anti-realist approaches as viable.
\end{quotation}

In the simple version of the argument, the authors consider preparing a quantum system (a qubit) in one of two quantum states: $|0\rangle$ or $|+\rangle = \frac{1}{\sqrt{2}}(|0\rangle + |1\rangle)$, where $|\langle 0|+\rangle| = \frac{1}{\sqrt{2}}$. Say $|0\rangle$ and $|1\rangle$ are eigenstates of $z$-spin, and $|+\rangle$ and $|-\rangle$ are eigenstates of $x$-spin. The supposition of the  theorem is that the epistemic states $|0\rangle$ and $|+\rangle$ overlap in the  ontic state space.  There is therefore a certain finite probability $q$ that preparing the system in either of these states will result in an ontic state $\lambda$ in the overlap region. If, now, two independent non-interacting systems are each prepared in either of the quantum states $|0\rangle$ or $|+\rangle$, there is a probability $q^{2}$ that the systems will be in ontic states $\lambda_{1}, \lambda_{2}$ that are both in the overlap region. It follows that the ontic state of the  2-qubit system (which the argument assumes is specified by the pair of local ontic states ($\lambda_{1}, \lambda_{2}$), because the systems are independent and non-interacting)\footnote{Note that it is not an assumption of the argument that $\lambda = (\lambda_{1}, \lambda_{2})$ is a common cause of any correlations between the two systems. That is, for measurements of observables $A,B$ on the 2-qubit system with outcomes $a$ for $A$ and $b$ for $B$, it is not required that $p(a,b|\lambda)  = p(a|\lambda)p(b|\lambda)$.} is compatible with the four quantum states: $|0\rangle\otimes |0\rangle, |0\rangle\otimes |+\rangle, |+\rangle\otimes |0\rangle,  |+\rangle\otimes |+\rangle$. 

Now consider measuring an observable $A$  on the 2-qubit system with entangled eigenstates:
\begin{eqnarray}
|0\rangle\otimes |1\rangle + |1\rangle\otimes |0\rangle \\
|0\rangle\otimes |-\rangle + |1\rangle\otimes |+\rangle \\
|+\rangle\otimes |1\rangle + |-\rangle\otimes |0\rangle \\
|+\rangle\otimes |-\rangle + |-\rangle\otimes |+\rangle
\end{eqnarray}

The outcome of the measurement should depend only on the ontic state,  which `screens off' future events from epistemic states (i.e., given the ontic state, an epistemic state provides no further information relevant to the occurrence of the event). But, according to quantum mechanics, the outcome corresponding to $|0\rangle\otimes |1\rangle + |1\rangle\otimes |0\rangle$ is impossible (i.e., has zero probability) if the initial   quantum state is  $|0\rangle\otimes |0\rangle$, since $|0\rangle\otimes |0\rangle$ is orthogonal to $|0\rangle\otimes |1\rangle + |1\rangle\otimes |0\rangle$. Similarly, $|0\rangle\otimes |+\rangle$ is orthogonal to $|0\rangle\otimes |-\rangle + |1\rangle\otimes |+\rangle$, $|+\rangle\otimes |0\rangle$ is orthogonal to $|+\rangle\otimes |1\rangle + |-\rangle\otimes |0\rangle$, and $|+\rangle\otimes |+\rangle$ is orthogonal to $|+\rangle\otimes |-\rangle + |-\rangle\otimes |+\rangle$. So each of the four possible outcomes of the measurement is impossible if the ontic state of the 2-qubit system is ($\lambda, \lambda'$). This is a contradiction, since one of the outcomes must occur, i.e., the sum of the probabilities of the four possible outcomes is 1.

The conclusion PBR draw is that the quantum state must be real, i.e., that the quantum state must be (at least part of) the ontic state. The argument rules out Leifer's third possibility, that quantum states are epistemic: distinct quantum states must correspond to physically distinct states of reality \cite[p. 476]{PBR2012}:
\begin{quote}
We have shown that the distributions for $|0\rangle$ and $|+\rangle$ cannot overlap. If the same can be shown for any pair of quantum states $|\psi_{0}\rangle$ and $|\psi_{1}\rangle$, then the quantum state can be inferred uniquely from $\lambda$. In this case, the quantum state is a physical property of the system.
\end{quote}
They proceed to prove this, from which it would seem that the only viable options for an interpretation that takes quantum mechanics as strictly true are Bohm's theory or the Everett interpretation, since Leifer's second possibility (`wave functions are epistemic, but there is no deeper underlying reality') is dismissed as instrumentalism. The general argument involves considering $n$ copies of the system, where $n$ depends on the angle between the states  $|\psi_{0}\rangle$ and $|\psi_{1}\rangle$ (the smaller the angle between the states, the larger the number $n$ of copies that needs to be considered).

The first thing to note here is that the PBR argument does not involve a locality assumption, so this result is different from Bell's related result \cite{BellEPR}. But there is an interesting relation with what Harvey Brown \cite{Brown1992} calls `Bell's other theorem': Bell's proof of a corollary to Gleason's theorem in \cite{Bellhv}, which anticipated the Kochen-Specker theorem \cite{KochenSpecker}.

Bell showed that if, in a Hilbert space of more than two dimensions, two nonorthogonal pure states are assigned 0 and 1, respectively, then if these states are close enough (i.e., if the angle between the rays is smaller than a certain angle), there exists a set of rays that are impossible to `color' noncontextually with $0$'s and $1$'s, consistently with the orthogonality constraints (where consistency requires that if a ray is assigned $0$, the orthogonal ray is assigned $1$, and conversely). 

Consider any two pure states in a Hilbert space of any dimension (including dimension 2), say $|\psi\rangle$ and $|\phi\rangle$. Suppose there is an ontic state $\lambda$ compatible with both of them. Represent this by assigning both states  $1$, where assigning $1$ to a quantum state \emph{qua} epistemic state indicates that the ontic state is in the support of the quantum state. 

Now consider $n$ noninteracting copies of the system and the two states:
\begin{eqnarray}
|\psi\rangle \otimes|\psi\rangle \otimes |\psi\rangle \cdots \\
|\phi\rangle \otimes|\phi\rangle \otimes |\phi\rangle \cdots 
\end{eqnarray}
As $n$ increases, the two states become more and more orthogonal (their scalar product gets smaller and smaller, which means that the angle between them gets larger and larger). It follows that, for sufficiently large $n$, there is some state $|\chi\rangle$ orthogonal to $|\psi\rangle \otimes|\psi\rangle \otimes |\psi\rangle \cdots$that is arbitrarily close to $|\phi\rangle \otimes|\phi\rangle \otimes |\phi\rangle \cdots$.  If $|\psi\rangle \otimes|\psi\rangle \otimes |\psi\rangle \cdots$ is assigned 1, the state $|\chi\rangle$ is assigned 0.

So if two nonorthogonal states are  both assigned $1$, and we take sufficiently many noninteracting copies of the system, there will exist two states in the tensor product Hilbert space that are closer than any given angle, such that one of the states is assigned 1 and the other 0. From Bell's result, it follows that there is a set of additional rays in this tensor product Hilbert space from which it is possible to derive a Kochen-Specker contradiction, i.e., it is impossible to assign these additional rays $0$'s and $1$'s noncontextually and also assign the two close rays $0$ and $1$ without violating orthogonality constraints. 

Of course, Bell's proof of the corollary to Gleason's theorem and the Kochen-Specker proof assume noncontextuality, that the same ray is assigned the same value in different orthogonal sets. What PBR have shown is that you can avoid the noncontextuality assumption and still get a contradiction.

If the two rays $|\psi\rangle \otimes|\psi\rangle \otimes |\psi\rangle \cdots$ and $|\phi\rangle \otimes|\phi\rangle \otimes |\phi\rangle \cdots$ are both assigned $1$, and $1$ is also assigned to all the other possible tensor combinations of $|\psi\rangle$'s and $|\phi\rangle$'s (all $n$-strings of $|\psi\rangle$'s and $|\phi\rangle$'s, for sufficiently large $n$) then---this is what PBR prove---there exists one (maximal) orthogonal set in the $2^{n}$-dimensional Hilbert space, i.e., a basis, that cannot be colored with $0$'s and $1$'s consistently with $1$'s assigned to the $2^{n}$ nonorthogonal rays (i.e., it is impossible to assign all but one of these basis rays 0 and also assign all the $2^{n}$ nonorthogonal rays  $1$ without violating orthogonality constraints). So noncontextuality does not arise. In effect, the noncontextuality assumption is avoided by considering many noninteracting copies of a system and assuming that each of the copies can be prepared in a given quantum state independently of the others, and that this also applies to ontic states, so that there are no `supercorrelations' at the ontic level and the ontic state of the composite many-copy system is  specified by the set of independent ontic states of the copies.

Bell rejected his own proof of the corollary to Gleason's theorem because he thought the noncontextuality assumption was not physically motivated \cite[p. 9]{Bellhv}: 
\begin{quote}
\ldots it relates in a nontrivial way the results of experiments which cannot be performed simultaneously; the dispersion free states need not have this property, it will suffice if the quantum mechanical averages over them do.
\end{quote}
His nonlocality theorem \cite{BellEPR} was an attempt to derive a similar conclusion from a physically motivated assumption. So one could say that the PBR result is an ingenious way of proving what Bell originally wanted to prove.

There is a similar result by Colbeck and Renner \cite[abstract]{ColbeckRenner2012}:
\begin{quote}
Here we show, based only on the assumption that measurement settings can be chosen freely, that a system's wave function is in one-to-one correspondence with its elements of reality. This also eliminates the possibility that it can be interpreted subjectively.
\end{quote}

The `free choice' assumption that measurement settings can be chosen freely is the assumption that a measurement setting can be chosen independently of any information that is in principle available before the choice of measurement setting, i.e.,  independently of any pre-existing value of any parameter, in any frame of reference.

The Colbeck-Renner investigation amounts to considering the following issue: Bell's nonlocality theorem \cite{BellEPR} established that  the correlations between certain measurement outcomes for a bipartite system in a certain pure entangled quantum state cannot be attributed to common causes in the common past of the two systems. A parameter $\lambda$ is a common cause of correlations if, conditional on $\lambda$, the correlations vanish. In other fords, the joint probabilities are conditionally statistically independent, conditional on the common cause. Conditional statistical independence is equivalent to the conjunction of two conditions: parameter independence and outcome independence.\footnote{The terminology is due to Abner Shimony \cite{Shimony1984}. For a proof of the equivalence here, see \cite[pp. 66--67]{Bubbook}.} Parameter independence (Bell locality) is the assumption that, conditional on $\lambda$, Alice's marginal probabilities are independent of Bob's measurement setting choices, and conversely. Outcome independence is the assumption that, conditional on $\lambda$, Alice's marginal probabilities are independent of Bob's measurement outcomes, and conversely. Parameter independence is well-motivated because, assuming access to $\lambda$, if this assumption were violated, then instantaneous signaling would be possible.\footnote{The no-signaling principle does not refer to any hidden variables or ontic states. Rather, the condition is that Alice's marginal probabilities, given the quantum state, are independent of Bob's measurement setting choices or, more generally, independent of anything Bob does. The no-signaling principle is parameter independence averaged over the hidden variables---an operational or phenomenal condition.}
 But outcome independence does not have the same plausibility. If parameter independence is violated, instantaneous signaling between Alice and Bob would be possible, given access to $\lambda$, because they choose the measurement settings. If outcome independence is violated, they cannot exploit the violation to signal instantaneously,  even given access to $\lambda$, because measurement outcomes are random and not under their control. They can choose the measurement settings, but not the measurement outcomes. So it is an interesting question to consider whether one can prove a `no go' theorem from parameter independence without assuming outcome independence: is an extension of quantum mechanics with improved predictive power possible, where the extended theory satisfies parameter independence but not necessarily outcome independence?Ê

Colbeck and Renner prove a result that answers the question negatively: no extension of quantum mechanics can improve predictability unless  the extension is such that,  for measurements on a bipartite system in a pure entangled state, parameter independence is violated: Alice's marginal probabilities in the extended theory would have to depend on Bob's measurement setting choices.ÊThis is a stronger version of Bell's nonlocality theorem. 

As we saw in \S 1 in the discussion of the measurement problem for a PR-box, if Bob has access to a parameter or ontic state $\lambda$ that completely specifies the output values for both of his possible inputs, or even specifies whether these output values are the same or different (a more coarse-grained specification), then either there is a violation of the no-signaling principle, or Alice's choice of input is not free but depends on the value of $\lambda$.  Since the alternative here depends on whether Bob obtains information about $\lambda$ before or after Alice's input, a violation of the no-signaling principle entails a violation of  Alice's freedom to choose her input, and conversely. 

A violation of the no-signaling principle,  given $\lambda$, is a violation of parameter independence. The `free choice' condition in the Colbeck-Renner theorem includes conditionalization with respect to the quantum state and ontic state $\lambda$, or the values of any hidden variables (since none of this is in the future light cone of  the choice of measurement setting).\footnote{In both the Pusey-Barrett-Rudolph analysis and the Colbeck-Renner analysis, nothing precludes the possibility that the quantum state could be part of the ontic state, or all of the ontic state. The `hidden variable' terminology suggests that the full ontic state is the quantum state together with additional hidden variables. A `no go' theorem for hidden variables therefore establishes that the quantum state is ontic, i.e., `the whole story.'} So a violation of the  `free choice' condition in the Colbeck-Renner theorem is equivalent to a violation of parameter independence.

We can see the significance of the Colbeck-Renner result by considering what one might call `Bohm's theory for a PR-box.' First note that a PR-box converts the truth value of the conjunction of the inputs (with $0$ corresponding to `false' and $1$ corresponding to `true') to the parity of the outputs.\footnote{`Outputs same' = parity $0$; `outputs different' = parity $1$.} So one could generate the correlations of a PR-box from a box with an internal stochastic mechanism that produces outputs for given inputs in a way that depends on the temporal order of the inputs:
\begin{description}
\item  if $x$ occurs before $y$ then, with equal probability, $a = 0$ and $b=x\cdot y$, or $a = 1$ and $b=x\cdot y \oplus 1$
\item  if $y$ occurs before $x$ then, with equal probability, $b = 0$ and $a=x\cdot y$, or $b = 1$ and $a=x\cdot y \oplus 1$
\item  if $x$ and $y$ occur simultaneously then, with equal probability, $a = 0$ and $b=x\cdot y$, or $a = 1$ and $b=x\cdot y \oplus 1$,  or $b = 0$ and $a=x\cdot y$, or $b = 1$ and $a=x\cdot y \oplus 1$
\end{description}

This produces the PR-box correlations $a\oplus b = x\cdot y$, and Alice's marginal probabillities are both $1/2$ independent of Bob's input, and similarly for Bob's marginals. So this box is non-signaling and phenomenally indistinguishable from a PR-box. Note that an output value, 0 or 1, is produced with probability $1/2$ irrespective of the remote input value, or whether there was any remote input value at all. 

There is, however, a big difference between the box as defined above and a PR-box, although the two boxes are empirically equivalent. In a PR-box, the temporal order of the inputs is irrelevant, and the correlations arise as a global feature of the statistics. In the box defined above, the temporal order of the inputs is relevant to the output values obtained, which requires the assumption of a preferred foliation in space-time.  Since the output values are separately defined as functionally related to the inputs, depending on the order of the inputs, and the correlations arise from these functional relations, one could construct a nonlocal deterministic hidden variable theory for the correlations as a causal explanation of the correlations in terms of the hidden variable  as a common cause---hence `Bohm's theory for a PR-box.'

Suppose the internal mechanism---the additional mathematical machinery in this extension of a PR-box---involves a register that keeps track of which input occurs first, or whether the inputs occur simultaneously, and a device with a hidden variable,  $\lambda$, uniformly distributed over $[0,1]$.  If  $x$ occurs before $y$, the device stores the input $x$ and produces an output $a = 0$ if  $\lambda < 1/2$ and $a = 1$ if  $\lambda \geq 1/2$. When the input $y$ occurs some time later, the device passes both inputs through an AND gate. If  $\lambda < 1/2$, the output of the AND gate is  transferred to Bob's output $b$. If $\lambda \geq 1/2$, then $1$ is added (mod $2$) to the output of the AND gate before it is transferred to Bob's output $b$. A similar sequence occurs if $y$ occurs before $x$, with $a$ and $b$ switched.  If the two inputs occur simultaneously, the device passes both inputs through the AND gate. If $\lambda < 1/4$, the device produces outputs $a = 0$ and $b = x\cdot y$; if  $1/4 \leq \lambda < 1/2$, the device produces outputs $a = 1$ and $b = x\cdot y \oplus 1$; if $1/2 \leq \lambda < 3/4$, the device produces outputs $b = 0$ and $a = x\cdot y$; if $3/4 \leq \lambda \leq 1$, the device produces outputs $b = 1$ and $a = x\cdot y \oplus 1$. Averaging over the hidden variables yields the PR-box correlations.

For a given value of the hidden variable $\lambda$, the Bohm box violates the no-signaling principle because Alice's input is revealed in Bob's output. That is, the Bohm box violates parameter independence. Suppose, for example, that Alice inputs $0$ or $1$ before Bob. If   $\lambda < 1/2$:
\begin{itemize}
\item[$\bullet$] $p(b=0| x = 0, y=1) = 1$ 
\item[$\bullet$] $p(b=1| x=1, y=1) =1$
\end{itemize}
Similarly,   if $\lambda \geq 1/2$ and Bob's input is $y = 1$, then Bob's output  is $1$ if Alice's input  is $0$, and 0 if Alice's input is $1$. So Alice and Bob could signal with a supply of Bohm boxes if they could control or measure the value of $\lambda$. Alternatively, if Bob obtains information about the value of $\lambda$ before Alice's input, there is a constraint on Alice's freedom of choice. Averaging   over $\lambda$ precludes the possibility of signaling and restores `free choice.' This is like Bohm's theory for quantum mechanics, which allows instantaneous signaling in principle, given sufficient control over  the hidden variables. In Bohm's theory, conflict with experience is avoided  because measurements cannot yield enough information about the hidden variables to allow signaling or violate `free choice,' assuming a certain equilibrium distribution for the hidden variables, which the theory guarantees will be maintained once achieved. 

A Bohm box illustrates the significance of the Colbeck-Renner result. In a Bohm box, the dynamics of the internal mechanism---the register and the device with the AND gate---would have to function instantaneously to simulate the correlations of a PR-box, which are assumed to be maintained for arbitrary separations of  the Alice and Bob inputs and outputs. The internal mechanism violates parameter independence, i.e., access to the internal mechanism would allow instantaneous signaling, or violate `free choice.' If there is no access in principle to the internal mechanism, then a Bohm box and a PR-box are empirically indistinguishable and nothing can rule out the possibility that a PR-box is really a Bohm box with an intrinsically hidden internal mechanism. A similar observation applies to Bohm's theory and quantum mechanics.   

For all we know, Bohm's theory might be true. But one might say the same for Lorentz's theory in relation to special relativity, insofar as it `saves the appearances.' Lorentz's theory provides a dynamical explanation of phenomena, such as length contraction, that are explained kinematically in special relativity in terms of the structure of Minkowski space-time. The theory does this at the expense of introducing motions relative to the aether---a preferred foliation in space-time---that are in principle unmeasurable, given the equations of motion of the theory. Similarly, Bohm's theory provides a dynamical explanation of quantum phenomena, such as the loss of information on measurement, which are explained kinematically in quantum mechanics in terms of the structure of Hilbert space, at the expense of introducing  the positions of the Bohmian particles, which are in principle unmeasurable more precisely than the Born distribution in the equilibrium theory, given the equations of motion of the particles.

\section{The Measurement Problem}
\label{sec:meas}

One could interpret the PBR result and the Colbeck-Renner result as showing that, at the fundamental level, systems are not characterized by an Einsteinian `being thus' or `So-sein.' 

The conclusion the authors draw is rather that the quantum state is ontic. For PBR, the quantum state could be part of the  full ontic state, as in Bohm's theory, or it could be the whole ontic state. For Colbeck-Renner, the possibility of adding anything to the ontic state is excluded if we assume parameter independence or, as they prefer to put it, `free choice.' The  only alternative possibility is a psi-epistemic view along the lines of Bohr's Copenhagen interpretation, or Fuchs' QBism, but this is dismissed as ultimately just old-fashioned instrumentalism, and the objection to instrumentalism as an interpretation of quantum mechanics is that it fails to address the measurement problem. For the instrumentalist, quantum mechanics is a theory about measurement outcomes, so how a measuring instrument produces definite outcomes remains outside the theoretical story. 

Here is how Bohm puts the measurement problem \cite[p. 583]{Bohm}:
\begin{quotation}
If the quantum theory is to be able to provide a complete description of everything that can happen in the world \ldots it should also be able to describe the process of observation itself in terms of the wave functions of the observing apparatus and those of the system under observation. Furthermore, in principle, it ought to be able to describe the human investigator as he looks at the observing apparatus and learns what the results of the experiment are, this time in terms of the wave functions of the various atoms that make up the investigator, as well as those of the observing apparatus and the system under observation. In other words, the quantum theory could not be regarded as a complete logical system unless it contained within it a prescription in principle for how all these problems were to be dealt with. 
\end{quotation}

The measurement problem is a \emph{consistency problem}. What we have to show is that the dynamics, which generally produces entanglement between two coupled systems, is consistent with the assumption that something definite or determinate happens in a measurement process. The basic question is whether it is consistent with the unitary dynamics to take the macroscopic measurement `pointer' or, in general, the macroworld as definite. The answer is `no,' if we accept an  interpretative principle sometimes referred to as the `eigenvalue-eigenstate link.' 

Dirac's states  the principle as follows \cite[pp. 46--47]{Dirac}\footnote{ One finds similar remarks in von Neumann \cite[p. 253]{Neumann} and in the EPR paper \cite{EPR}. The EPR argument is formulated as a \emph{reductio} for the principle: EPR show that it follows from the principle, together with certain realist assumptions, that quantum mechanics is incomplete}:
\begin{quote}
The expression that an observable `has a particular value' for a particular state is permissible in quantum mechanics in the special case when a measurement of the observable is certain to lead to the particular value, so that the state is an eigenstate of the observable. \ldots In the general case we cannot speak of an observable having a value for a particular state, but we can speak of its having an average value for the state. We can go further and speak of the probability of its having any specified value for the state, meaning the probability of this specified value being obtained when one makes a measurement of the observable.
\end{quote}

Dirac's principle, together with the linearity of the unitary dynamics of quantum mechanics, leads to the measurement problem: the definiteness or determinateness of a particular measurement outcome is inconsistent with the entangled state that linearity requires at the end of a measurement. Typically, the measuring instrument is a macroscopic system, like Schr\"{o}dinger's cat, where the cat states $\ket{\mbox{alive}}$ and $\ket{\mbox{dead}}$ are correlated with eigenstates of a microsystem in the final entangled state at the end of the dynamical interaction. Here the macroscopic cat states act as measurement pointer states for the eigenvalues of the measured observable of the microsystem. Schr\"{o}dinger's point was that, while we might find the indefiniteness of a microsystem in a superposition acceptable, we surely would not want to say that the cat is indefinite with respect to being alive or dead when it is entangled with a microsystem, i.e., when the composite cat-microsystem is in a superposition of alive and dead cat-states.

In quantum mechanics, the transition in a measurement process from a state
$\ket{s}$, in which the event $s$ is definite according to the eigenvalue-eigenstate link, to a different nonorthogonal state $\ket{t}$, in which the event $t$ is definite, with probability  $|\langle t|s\rangle|^{2}$   is a stochastic transition that is not described by the unitary dynamics of the theory. Since the event $s$ was definite before the measurement and is now, in the state $\ket{t}$  after the occurrence of the measurement outcome $t$, indefinite, so the probability of finding $t$ in a measurement is less than 1, there is a loss of information on measurement or, as Bohr put it, an `irreducible and uncontrollable' measurement disturbance. As we saw in \S 1, this is a general feature of any no-signaling theoretical explanation of correlations that are outside the local correlation polytope. 

Just as Lorentz contraction is a physically real phenomenon
explained relativistically as a kinematic effect of motion in a
non-Newtonian space-time structure, so the change arising in
quantum conditionalization that involves a real loss of information should be understood as a kinematic effect of \emph{any} process
of gaining information of the relevant sort in the non-Boolean possibility structure
of Hilbert space, considered as a kinematic framework for an indeterministic physics  (irrespective of the dynamical processes involved in the
measurement process).   The definite occurrence of a particular event is constrained
by the kinematic probabilistic correlations represented by the subspace
structure of Hilbert space, and only by these correlations---it is otherwise
free.

The eigenvalue-eigenstate link is an interpretative principle, a \emph{stipulation} about when an observable `has a particular value,' that is not  required by the kinematic structure of quantum mechanics. Alternative stipulations are possible. In particular, no contradiction is involved in stipulating that some particular `preferred' observable, $R$, is always definite (as long as we don't stipulate other noncommuting observables as also definite). 

The following lattice-theoretic theorem \cite{BubClifton} (see \cite{HalvorsonClifton1999} for a $C^{*}$-algebraic generalization, and \cite{Nakayama2008} for a topos-theoretic generalization) characterizes all possible stipulations, subject to some minimal constraints: If $R$ is a preferred observable in this sense (stipulated as always definite, i.e., as always having a determinate value), and $e$ is the ray representing a pure quantum state, there is a unique maximal sublattice $\mathcal{L}(R,e)$ in the lattice of Hilbert space subspaces representing quantum propositions, such that:
\begin{itemize}
\item[$\bullet$] elements of  $\mathcal{L}(R,e)$ can be simultaneously definite in the state $e$
\item[$\bullet$]  $\mathcal{L}(R,e)$ contains the spectral projections of the preferred observable $R$
\item[$\bullet$]  $\mathcal{L}(R,e)$ is defined by $R$ and $e$ alone (i.e., $L(R,e)$ is invariant under Hilbert space lattice automorphisms that preserve $R$ and $e$)
\end{itemize}
Although  $\mathcal{L}(R,e)$ is not a Boolean algebra, in general, there are sufficiently many\footnote{But not sufficiently many to distinguish every pair of distinct elements, as there is in a Boolean algebra.}  2-valued homomorphisms (corresponding to truth-value assignments) on  $\mathcal{L}(R,e)$ so that a probability measure can be defined on  $\mathcal{L}(R,e)$, such that the probability of an element $a$ in  $\mathcal{L}(R,e)$ given by the quantum state $e$ is just the measure of the set of 2-valued homomorphisms that assign 1 to $a$.

The elements in $\mathcal{L}(R,e)$ are obtained by projecting $e$ onto the eigenspaces of $R$. These projections, together with all the rays in the subspace orthogonal to the span of the projections, generate $\mathcal{L}(R,e)$. In the case of Schr\"{o}dinger's cat, if we take $R$ as the cat-observable with eigenstates $|\mbox{alive}\rangle$ and $|\mbox{dead}\rangle$, then it follows that the cat is definitely alive or definitely dead in the entangled state after the interaction with the microsystem, and the microsystem property correlated with the cat being definitely alive or the cat being definitely dead is also definite in the final entangled state.

The eigenvalue-eigenstate link, or Dirac's principle, amounts to the stipulation that the preferred observable $R$ is the identity $I$, which has the whole Hilbert space and the null space as the two eigenspaces. There is a sense in which the eigenstate-eigenvalue link is preserved by the new stipulation. If a system $S$ is not entangled with the decoherence pointer $R$, so that the total state is a product state $|s\rangle\otimes \cdots$, where $|s\rangle$ is the state of $S$, then the decoherence `pointer'  takes the form $I\otimes R$, where $I$ is the identity in the Hilbert space of $S$, and the definite $S$-properties according to the above theorem are as specified by the eigenstate-eigenvalue link.

If we take $R$ as the decoherence `pointer' selected by environmental decoherence, then it follows that the macroworld is always definite because of  the nature of the decoherence interaction coupling environmental degrees of freedom to macroworld degrees of freedom (a contingent feature of the quantum dynamics), and it follows from the theorem that features of the microworld correlated with $R$ are definite. In other words, decoherence guarantees the continued definiteness or persistent objectivity of the macroworld, if we stipulate that $R$ is the decoherence `pointer.'

The argument  here is not the usual argument that decoherence provides a dynamical explanation of how an indefinite quantity becomes definite in a measurement process---Bell \cite{BellCH} has aptly criticized this argument as a `for all practical purposes' (FAPP) solution to the measurement problem. Rather, the claim is that we can take the decoherence pointer as definite \emph{by stipulation}, and that decoherence then guarantees the objectivity of the macroworld, which resolves the measurement problem without resorting to Copenhagen or neo-Copenhagen instrumentalism. In effect, the theorem shows that you can take the different histories associated with the branching structure in an Everettian interpretation of quantum mechanics as possible histories, one of which is the actual history. So there is no need to invoke the Everettian multiverse to solve the measurement problem. 

The consistency argument does not, of course, explain how an individual measurement outcome is selected out of all possible outcomes. This is a feature of any no-signaling theoretical explanation of correlations that lie outside the local correlation polytope.  In a simplex theory, it is possible to give a dynamical account of how an individual measurement outcome is selected as a dynamical transition from one vertex of the simplex to another.  For correlations that cannot be simulated in a simplex theory, there is no analogous dynamical account, because any such account would be inconsistent with the no-signaling principle. The appropriate answer to the question about how an individual measurement outcome is selected in a quantum measurement process---how `the transition from the possible to the actual' comes about---is that this occurs `freely,' as a truly random event, consistent with the probabilistic correlations of the quantum convex set. There is nothing more that needs to be said or can be said about how the trick is done.

\section{Concluding Remarks}
\label{sec:conc}

We know how to solve the measurement problem: Bohm's theory is a solution, the Everett interpretation is a solution, the Ghirardi-Rimini-Weber theory \cite{GhirardiSEP} is a rival theory  that avoids the measurement problem. The debate continues because there is nothing like a general consensus that any of these proposals are getting it right. Einstein commented in a letter to Max Born \cite[p. 192]{Born} that Bohm's theory `seems too cheap to me.'  One might say the same about all these solutions: they explain away the irreducible indeterminism of quantum mechanics, rather than providing a conceptual framework for thinking about a universe in which, to put it somewhat anthropomorphically, a particle is free to choose its own response to a measurement, subject only to probabilistic constraints, which might be nonlocal. 

The transition from classical to quantum mechanics rests on the recognition that the structure of information is not what we thought it was: there are probabilistic correlations that lie outside the polytope of local correlations, and so cannot be simulated by `shared randomness' or common causes. The `no go' theorems tell us that we can't  explain these correlations in terms of a classical simplex theory by supposing that something has been left out of the story. The nonclassical features of quantum mechanics, including the irreducible information loss on measurement, are generic features of any no-signaling theoretical explanation of correlations that lie outside the local correlation polytope. Fundamentally, the conceptual problem is how to make sense of this---how to understand a nonlocal box theory. A large part of the difficulty is removed by seeing that the inconsistency in reconciling the entangled state at the end of a quantum measurement process with the definiteness of the macroscopic pointer reading and the definiteness of the correlated value of the measured micro-observable is only apparent and depends on a stipulation that is not required by the structure of the quantum possibility space. Replacing this stipulation by an alternative consistent stipulation resolves the problem.

section*{acknowledgements}
Informative discussions with Allen Stairs and Tony Sudbury are gratefully acknowledged. My research is supported by the Institute for Physical Science and Technology at the University of Maryland. This publication was made possible through the support of a grant from the John Templeton Foundation. The opinions expressed in this publication are those of the author and do not necessarily reflect the views of the John Templeton Foundation.

\bibliographystyle{plain}
\bibliography{qcmp.bib}

\end{document}